\documentclass[smallextended]{svjour3}
\usepackage{graphicx}
\usepackage{amssymb}
\newcommand{\beq}{\begin{eqnarray}}
\newcommand{\eeq}{\end{eqnarray}}

\journalname{Low Temperature Physics}
\begin{document}
\title{Small-Angle Neutron Scattering and Magnetization Study of HoNi$_{2}$B$_{2}$C}
\author{M. Ramazanoglu \and
M. Laver \and
A. Yagmurcu \and
E.-M. Choi \and
S.-I. Lee \and
A. Knigavko \and
B.D. Gaulin
}
\institute{M. Ramazanoglu \at
Department of Physics and Astronomy, McMaster University, Hamilton, Ontario, L8S 4M1, Canada\\
Faculty of Engineering, Istanbul Technical University, 34469, Maslak, Istanbul, Turkey
\and
M. Laver \at
Department of Metallurgy and Materials, University of Birmingham, Edgbaston, Birmingham, B15 2TT, United Kingdom
\and
Abdulhamit Yagmurcu \at
Ministery of Development , Ankara, 06100, Turkey 
\and
E.-M. Choi \and S.-I. Lee \at
National Creative Research Initiative Center for Superconductivity and Department of Physics,
Pohang University of Science and Technology, Pohang 790-784, Republic of Korea
\and
A. Knigavko \at
Department of Physics, Brock University, St. Catharines, Ontario, L2S 3A1, Canada
\and
B.D. Gaulin \at
Department of Physics and Astronomy, McMaster University, Hamilton, Ontario, L8S 4M1, Canada\\
Canadian Institute for Advanced Research, 180 Dundas St. W.,Toronto, Ontario, M5G 1Z8, Canada
}
\date{Received: August 6, 2013}
\maketitle

\begin{abstract}
The superconducting and magnetic properties of HoNi$_{2}$B$_{2}$C single crystals are investigated through transport,
magnetometry and small-angle neutron scattering measurements.
In the magnetic phases that enter below the superconducting critical temperature,
the small-angle neutron scattering data uncover networks of magnetic surfaces.
These likely originate from uncompensated moments e.g.\ at domain walls pinned to crystallographic grain boundaries.
The field and temperature dependent behaviour appears consistent with the metamagnetic transitions reported in earlier works.
\end{abstract}

\section{Introduction}

The rare-earth nickel borocarbide compounds $R$Ni$_{2}$B$_{2}$C exhibit superconductivity even when the rare-earth element $R$
is a magnetic ion~\cite{Kreyssig,Lynn1,Gupta,Niewa,Muller}.
Due to the possibility of studying the interplay between superconductivity and magnetism, these materials have remained at the forefront of condensed 
matter physics research for over a decade following their discovery~\cite{Cava,Siegrist}.
For the non-magnetic ions $R = $Y and $R = $Lu, the superconducting critical temperatures
are $T_c = 15$\,K and $T_c = 6$\,K respectively~\cite{Lynn1,Siegrist,Rathnayaka}.
For the magnetic ions $R = $Ho, Tm or Er, N\'{e}el states enter at temperatures below $T_{c}$,
whereas for $R = $Dy superconductivity is stable only within
the antiferromagnetic region~\cite{Kreyssig,Lynn1,Hill,Dewhurst,Detlefs1,Naidyuk1,Naidyuk2,Eisaki,Sierks,Detlefs2}.
All the rare-earth nickel borocarbides share a nominally tetragonal crystal structure,
though orthorhombic distortions can appear due to magnetoelastic effects~\cite{Kreyssig,Muller}.
The unit cell is formed of alternating $R$C and Ni$_{2}$B$_{2}$ layers,
with superconductivity understood to originate in the latter~\cite{Lynn1,Gupta,Niewa,Muller,Siegrist}.

In HoNi$_{2}$B$_{2}$C, the superconductivity appears at $T_c \sim 9$\,K.
Subsequently, a cascade of different magnetic structures within a narrow range of temperatures 6.0\,K to 5.2\,K
results in a near-reentrant behaviour of the superconducting phase\cite{Hill,Dewhurst,Detlefs1,Canfield1,Canfield2}.
The interplay between magnetism and superconductivity in HoNi$_2$B$_2$C has been succinctly probed
by Bitter decoration measurements by Vinnikov et al.\cite{Vinnikov},
where it was demonstrated that magnetic domain boundaries strongly pin superconducting vortices,
at least in the commensurate magnetic phase at low temperatures ($T < 5.2$\,K).
Similar pinning to magnetic domain boundaries was also observed in ErNi$_2$B$_2$C\cite{Vinnikov}.
Intriguingly, the antiferromagnetic state for $T < 6$\,K in ErNi$_2$B$_2$C develops a ferromagnetic component below 2.3\,K~{\cite{Lynn2}}
raising the possibility of a subsequent spontaneous formation of superconducting vortices~\cite{Lynn2}.
It would be intriguing to see if similar effects should occur at the metamagnetic transitions in HoNi$_2$B$_2$C.
This article is the first report of a small-angle neutron scattering (SANS) study on this material.
At zero and low applied fields we find sharply increasing SANS as the samples are cooled through $T \sim 5$\,K.
We associate this signal to scattering from magnetic surfaces (i.e.\ domain walls) in the low temperature collinear magnetic phases.

This article is continued as follows: in the following section (Sec.~\ref{experimental}) we outline the materials and methods;
in the Results section (Sec.~\ref{results}) we report the results of our resistivity (Sec.~\ref{resistivity}),
magnetization (Sec.~\ref{magnetizationsection}) and SANS (Sec.~\ref{SANSsection}) measurements.
A short summary and discussion is provided in the concluding section (Sec.~\ref{conclusion}).

\section{Experimental details}
\label{experimental}

Single crystals of HoNi$_{2}$B$_{2}$C were grown using a slow-cooling flux method with isotropic enrichment of $^{11}$B in order to
reduce the neutron absorption in our SANS experiments.
The details can be found elsewhere \cite{Mun}.
Magnetization and transport measurements were performed in the physics laboratories at McMaster University using 
a Quantum Design Magnetic Properties Measurement System and a Quantum Design Physical Properties Measurement System respectively.
Resistance was measured by the standard four-point probe technique.
The SANS experiments were carried out at the NIST Center for Neutron Research using the NG3-SANS and NG7-SANS instruments~\cite{Glinka1998}.
In a typical setup, $\approx 9$\,\AA\ wavelength neutrons were used
and the small-angle scattering detected with a 2D area detector placed $\sim 13.7$\,m away from the sample.
Two HoNi$_{2}$B$_{2}$C single crystals (8x5x1 and 5x4x1\,mm${^3}$) were co-aligned on an Al sample holder by X-ray Laue diffraction.
The holder was mounted into a superconducting cryomagnet so that the samples' \textbf{c}-axes, applied field (\textbf{H})
and neutron beam directions were all initially coincident,
with the $[1\bar{1}0]$ and $[110]$ crystallographic axes were aligned horizontal and vertical respectively.
The field direction \textbf{H} of the cryomagnet was initially aligned relative to the neutron beam using the vortex lattice in a Nb single crystal.
The HoNi$_2$B$_2$C scattering measurements detailed herein were carried out with \textbf{c} turned away by $\sim 45^\circ$,
so that \textbf{H} and the neutron beam direction were $\sim 45^\circ$ to \textbf{c} in the (110) plane.
Datasets as a function of temperature were collected by applying the desired field at high temperatures $> T_c$,
cooling to base temperature ($\lesssim 3$\,K) and then measuring at selected temperatures on warming.

\section{Results}
\label{results}
\subsection{Resistivity}
\label{resistivity}
\begin{figure}
\includegraphics[scale=.5]{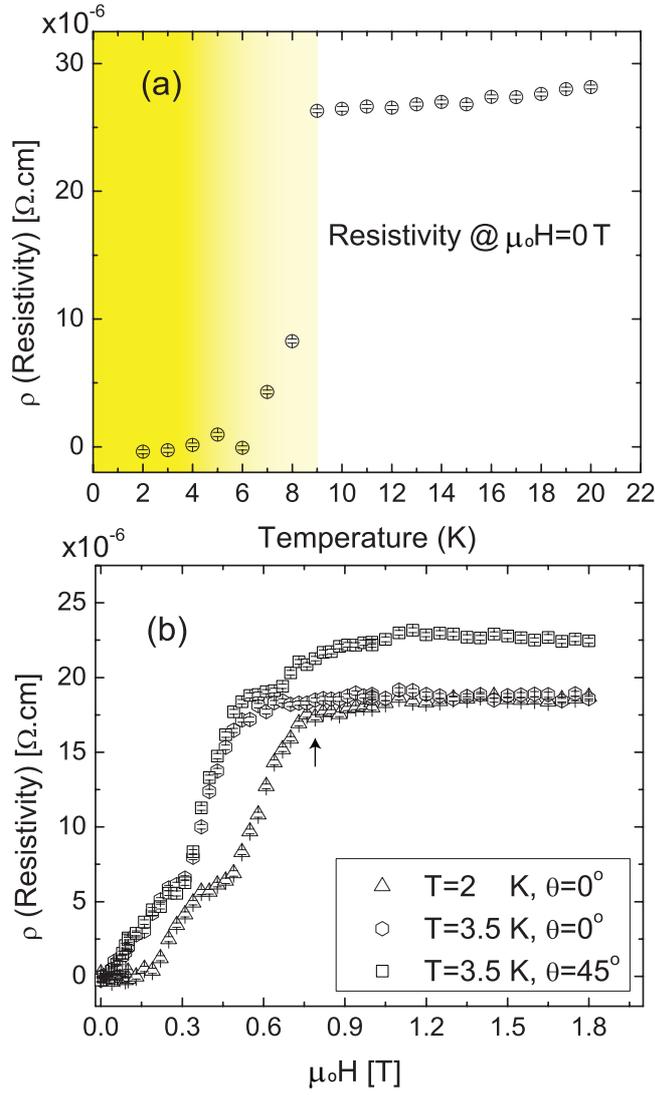}
\caption {\label{Resistance} Resistivity $\rho$ measurements 
as functions of temperature $T$ and of magnetic field $H$.
(a) shows $\rho(T)$ at zero field; (b) shows $\rho(H)$ at $T=2$\,K and $T=3.5$\,K, with fields \textbf{H}
applied parallel to \textbf{c} and also at 45$^\circ$ to \textbf{c} in the (110) plane.
The arrow shows the point for H$_{c2}$(T=2K) of the previous study~\cite{Dewhurst}.
In (a) the yellow background shading highlights the superconducting region.
}
\end{figure}

Figure~\ref{Resistance}(a) shows a plot of electrical resistivity $\rho$ as a function of temperature at zero field.
The current created by the four probe measurement lies in the (ab) plane of the crystal. 
As can be seen in this figure, the onset of superconductivity appears at $T_c \sim 9$\,K.
Overall metallic behaviour can also be seen for T$>$T$_{c}$ in the limited temperature range probed.
In Fig.~\ref{Resistance}(b), $\rho$ at low temperatures is shown as a function of applied field \textbf{H}
in the two directions, $0^\circ$ and $\sim 45^\circ$ from the crystallographic \textbf{c}-axis, for direct comparison with the SANS measurements.
At $T=2$\,K and \textbf{H}~$\parallel$~\textbf{c}, the upper critical field is found at $\mu_0 H_{c2}(T=2$\,K$)\sim 0.7$\,T,
in agreement with previous works~\cite{Dewhurst,Krug}.
The coherence length at $T=2$\,K from this value is $\xi\simeq 200${\AA}.
A kink at lower fields $\sim 0.4$\,T can also be observed in this figure.
We note that the slope of $\rho(H)$ steepens with increasing field through the kink,
implying that the superconductivity abruptly weakens.
This can be associated with increased fluctuations arising from the lower $T_{c2}(H)$ of the reentrant region.

Similar kinks are observed at $\approx 0.35$\,T in the $T=3.5$\,K datasets.
For \textbf{H}~$\parallel$~\textbf{c},
we see that $\mu_0 H_{c2}(T=3.5$\,K$)\sim 0.5$\,T in agreement with the expected form of $H_{c2}(T)$~\cite{Dewhurst}.
A slight decrease in $H_{c2}$ is anticipated on rotation of \textbf{H} from \textbf{c} towards the tetragonal basal plane~\cite{Krug},
and indeed a small leftwards shift of $\rho(H)$ in the upper critical field region may be discerned
going from \textbf{H}~$\parallel$~\textbf{c} to \textbf{H} applied $45^\circ$ from \textbf{c}.
In the latter, the $\rho(H)$ further rises into a broad maximum at fields above $H_{c2}$.
This may be a consequence of the metamagnetic transitions to be discussed in Sec.~\ref{SANSsection};
similar features in the resistivity have previously been reported for fields applied in the basal plane~\cite{Krug}.

\subsection{Magnetization}
\label{magnetizationsection}
\begin{figure}
\includegraphics[scale=.4]{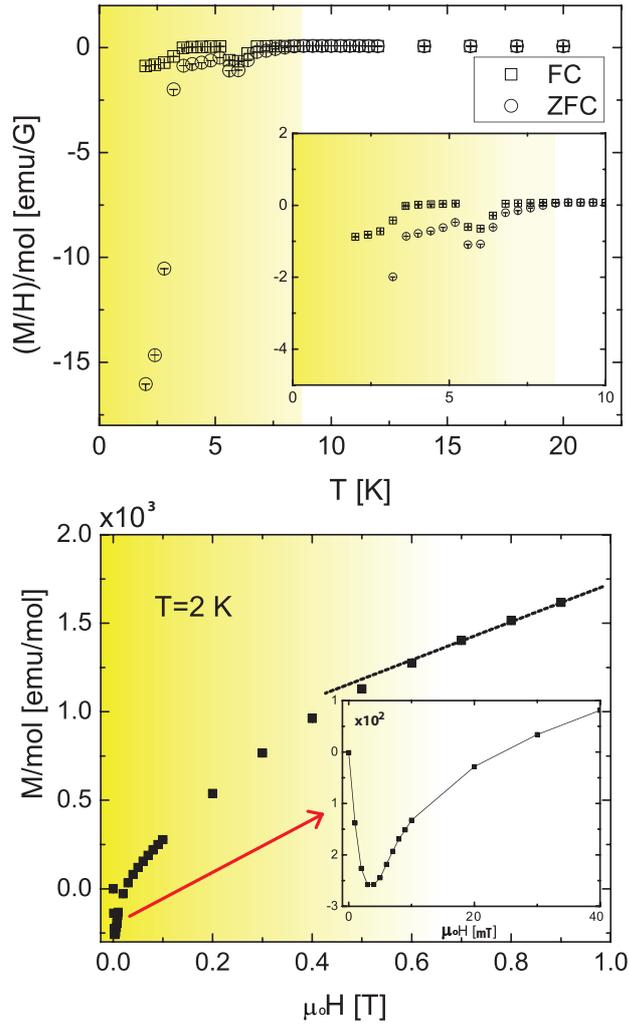}
\caption {\label{Magnetization} Magnetometry $M$ as functions of temperature $T$ and of magnetic field $H$.
(a) The temperature dependence of $M/H$, measured after cooling in zero field (ZFC) and field-cooling (FC)
using a small applied field of 1\,mT.
(b) $M(H)$ at $T=2$\,K. The dashed line indicates the linear paramagnetic response of a component of the Ho$^{3+}$ moments at high fields.
The inset in each panel shows a magnified region of the main figure.
In these measurements, the magnetic field is applied parallel to the \textbf{c} axis.
For clarity, yellow background shading has been added to highlight the superconducting region.
}
\end{figure}

Figure~\ref{Magnetization} illustrates our magnetometry data.
In Fig.~\ref{Magnetization}(a), the zero field cooled (ZFC) and field cooled (FC) magnetic responses in a small applied field of 1\,mT
are plotted as a function of temperature.
The reentrant superconductivity manifests itself as a reduced superconducting response in the region $T=5$\,K to 6\,K.
In Fig.~2(b) we show the magnetization at 2\,K as a function of increasing field applied parallel to \textbf{c} after cooling in zero field.
The overall signal is dominated by the linear paramagnetic response of a component of the ($\approx 10 \mu_B$) Ho$^{3+}$ moments,
but one can estimate the value of the upper critical field from the point of departure from the linear normal state behaviour
illustrated by the dashed line;
one then obtains $\mu_0 H_{c2}\sim 0.7$\,T in agreement with our resistivity measurements [Fig.~\ref{Resistance}(b)] and
previous Hall probe results~\cite{Dewhurst}.
The $M(H)$ data in Fig.~\ref{Magnetization}(b) can also be used to estimate the lower critical field value as $\mu_0 H_{c1}\sim 4$\,mT at $T=2$\,K,
though this estimate does not take into account sample demagnetization or Ho$^{3+}$ moment alignment effects.

In Fig.~2(b) we measure a linear paramagnetic response of 1750\,emu\,mol$^{-1}$T$^{-1}$, or 0.31\,$\mu_B$\,T$^{-1}$ per Ho$^{3+}$ ion.
It is interesting to compare this with previously published values.
Ref.~\cite{SchneiderThesis} reports 0.3$\mu_B$\,T$^{-1}/$ion in agreement with our measurement.
In contrast, a smaller value of 0.1$\mu_B$\,T$^{-1}/$ion is reported in Ref.~\cite{Canfield2}.
It should be noted that all works report comparable $M$ values at the metamagnetic transitions
present for \textbf{H}~$\perp$~\textbf{c}; this is expected for an intensive quantity like magnetization.
The possible non-intensive nature of $M$ for \textbf{H}~$\parallel$~\textbf{c}
suggests that its origin may not be an intrinsic property of the bulk of samples.
In the next section,
we moot the possibility that this magnetization arises from more easily polarisable Ho$^{3+}$ moments at crystallographic grain boundaries
and dislocations.

\subsection{Small-angle neutron scattering (SANS)}
\label{SANSsection}
\begin{figure}
\includegraphics[width=3.2in]{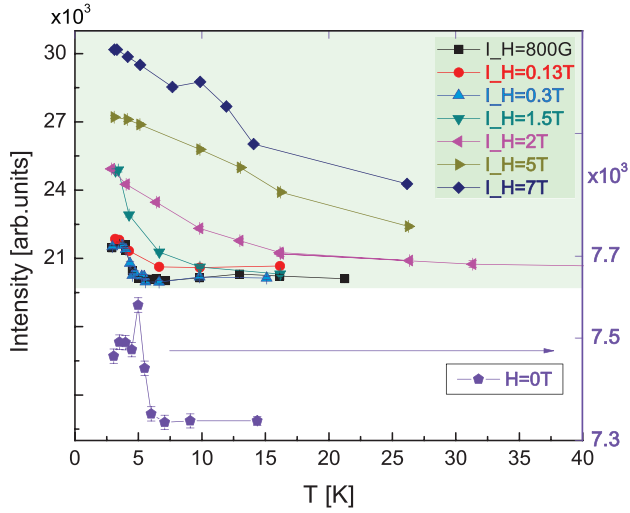}
\caption {\label{SANS} Integrated detector counts, summed over the range of ($\sim 3 \leq q \leq 14$)x10$^{-3}$\,\AA$^{-1}$ as a 
function of temperature at various magnetic fields applied 45$^\circ$ from \textbf{c} in the (110) plane.
The zero field data collected in a separate experimental setup have their own intensity scale (right).
}
\end{figure}

\begin{figure}
\includegraphics[width=3.2in]{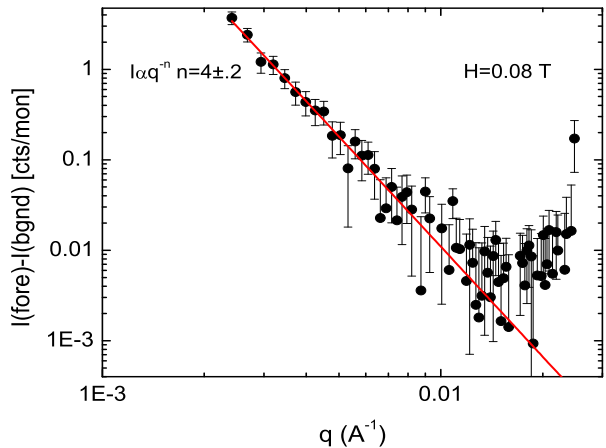}
\caption {\label{Ivsq} Intensity $I$ as a function of scattering vector $q$ at low temperatures and $\mu_0 H=0.08$\,T.
A high temperature background has been subtracted.
The red line shows the power-law behaviour of $I\propto q^{-n}$ with the fitted $n=4.0 \pm 0.2$.}
\end{figure}

We now turn to the main results of this manuscript.
Figures~\ref{SANS} and \ref{Ivsq} summaries the dependences of the small-angle neutron scattering (SANS) on temperature, applied field
and scattering vector $q$.
As a function of applied field, HoNi$_2$B$_2$C exhibits a series of metamagnetic transitions with magnetic structures depending
on the field direction \cite{Rathnayaka,Dewhurst,Detlefs1,Canfield1,Canfield2,Detlefs2,Krutzler,Krug,Campbell,Detlefs3}.
At low temperatures $T \lesssim 5$\,K with \textbf{H} applied along the $[110]$ easy axis, the magnetic structures
are constructed from ferromagnetic sheets, with each sheet being a basal plane,
within which all Ho$^{3+}$ moments co-align along one of the four $\langle 110 \rangle$ directions~\cite{Canfield2}.
The zero-field structure consists of antiferromagnetically coupled basal plane sheets,
denoted by $\uparrow \downarrow$.
At the first metamagnetic transition at 0.41\,T at 2\,K, a jump in the magnetization signals that two thirds of the basal
plane sheets align along \textbf{H} $\parallel [110]$ and only one along $[\bar{1}\bar{1}0]$;
this is denoted $\uparrow \uparrow \downarrow$.
Subsequently this becomes the high-field $\uparrow \uparrow \uparrow$ phase
with another magnetization jump at the second metamagnetic transition at 1.07\,T at 2\,K.
Fields applied away from $[110]$ in the basal plane yield a second intermediate field phase $\uparrow \uparrow \rightarrow $.
The $\rightarrow$ indicates that the magnetic structure is non-collinear;
it has a wave vector $\approx$($\frac47$~0~0)~r.l.u.\ \cite{Campbell,Detlefs3}.
A recent torque magnetometry study~\cite{Rathnayaka} confirms that this phase does not appear for \textbf{H} applied within $1^\circ$ of [110],
in support of the results of mean-field theory accounting for the crystalline electric field and
Ruderman-Kittel-Kasuya-Yosida exchange interaction~\cite{Amici}.


For fields applied along the hard \textbf{c}~axis, there are no metamagnetic transitions over this field regime.
Therefore, for our orientation of field direction $\approx 45^\circ$ from \textbf{c} in the (110) plane,
we anticipate two clear metamagnetic transitions;
the first from $\uparrow \downarrow$ to $\uparrow \uparrow \downarrow$ at $0.41 / \cos 45^\circ = 0.58$\,T 
and the second to $\uparrow \uparrow \uparrow$ at 1.51\,T.

Fig.~\ref{SANS} shows the SANS intensity integrated over the 2D detector
(this corresponds to a $q$-range of $(3 \leq q \leq 14) \times 10^{-3}$\,\AA$^{-1}$).
We see there are indeed three different field regimes that correspond to the expected metamagnetic transitions.
First, at low fields $< 0.58$\,T and including zero field, the SANS temperature dependences share a common form,
with intensity at low temperatures entering abruptly, rather like an order parameter with a transition temperature of $\approx 5$\,K.
Then, at an intermediate applied field of 1.5\,T, the magnitude of this order-parameter-like response appears somewhat increased
compared to the low field regime.
Finally at high fields $> 1.51$\,T the SANS intensity rises smoothly with falling temperature,
and also increases with increasing field.

To understand the origin of the SANS intensity appearing at low temperatures,
we first consider the possibility that the SANS signal at fields $< H_{c2}$ arises from superconducting vortices in the sample.
To check for sharp Bragg reflections from a well-defined superconducting vortex lattice,
at selected low applied fields we also collected data as the samples were rocked through the expected Bragg angles.
The field direction of the cryomagnet was initially aligned relative to the neutron beam using the vortex lattice in a Nb single crystal.
No angular dependence was observed in the rocking scans on our HoNi$_2$B$_2$C samples, precluding a well-defined vortex lattice.
At zero field, a SANS signal is observed even when no superconducting vortices are present
(no spontaneously forming vortices are anticipated because the magnetic ground state is antiferromagnetic).
Hence superconducting vortices, even in a disordered ensemble, cannot account for the observed SANS response.

Next we turn to the possibility that temperature-induced SANS is generated by multiple Bragg scattering
between slightly misoriented crystallites in the samples.
This possibility would be completely excluded by working at neutron wavelengths $\lambda$ above the Bragg cutoff $\lambda > 2 d_{max}$
where $d_{max}$ is the maximum $d$-spacing of the Bragg diffraction planes.
For HoNi$_2$B$_2$C, $d_{max} = 10.53$\,\AA, corresponding to the $(0 0 1)$ magnetic Bragg reflection that appears at $T < 5.2$\,K
in the commensurate N{\'e}el state.
Unfortunately working at $\lambda > 21$\,\AA\ is infeasible due to the lack of flux at these wavelengths at
even the best of today's neutron facilities. Instead we used $\lambda \approx 9$\,\AA.
Nonetheless we may defenestrate multiple Bragg scattering since this usually manifests itself
as isolated spots on the small-angle detector that typically change rapidly
with rotations of the sample or with neutron wavelengths.
Here, in contrast, we observe a smooth SANS profile (c.f.\ Fig.~\ref{Ivsq}) that is insensitive to small changes in angle or wavelength.

Fig.~\ref{Ivsq} shows the typical scattering vector $q=|$\textbf{q}$|$ dependence of the SANS intensity $I$ induced at low temperatures.
In this figure we plot the behaviour at 0.08\,T (high temperature $> 9$\,K backgrounds have been subtracted) but
a similar behaviour for the induced SANS intensity $I$ is observed at all fields.
The data fit well to a Porod law $I \propto q^{-n}$ with the fitted exponent $n \sim 4$ at all fields.
At 0.08\,T, for example, we find $n = 4.0 \pm 0.2$.
The atomic form factor $F_A(q)$ of the Ho$^{3+}$ moments is essentially constant at low $q$,
and the SANS profile probes directly how these moments arrange.
The Porod behaviour implies a network of interfaces or surfaces, with the $n \sim 4$ signifying that these interfaces are smooth.
Our experiments do not probe sufficiently small scattering vectors to access the Guinier regime of the $I(q)$ profile
and we therefore deduce a lower bound $D \gg 1300$\,\AA\ for the characteristic size $D$ of the network.

In the high field region, we find that the anisotropy of the temperature-induced SANS is very similar to that of the crystallographic background,
as characterized by the angular dependence of \textbf{q} measured over the 2D SANS detector.
The same can be said for the low field region, although here there are fewer counts with which to make this evaluation.
Like the temperature-induced foreground SANS intensity, the crystallographic background exhibits a Porod law $\propto |q|^{-4}$,
indicative of grain boundaries and dislocation surfaces.
It is likely that the temperature-induced SANS also originates at these crystal grain boundaries and dislocations.
In the antiferromagnetic $\uparrow \downarrow$ region at zero and low fields,
we conjecture that the grain boundaries and dislocations serve as pinning sites for antiferromagnetic domain walls
that carry a net magnetic moment.
Due to the crystal electric field, at high fields the Ho$^{3+}$ moments align to the [110] direction that is nearest to the applied field \textbf{H};
we postulate that additional small-angle scattering appears at decreasing temperatures from interfacial moments at the grain boundaries and dislocations
that are susceptible to align directly along \textbf{H} instead of along $[110]$.
As mooted in Sec.~\ref{magnetizationsection}, these free moments at extrinsic interfaces may also account
for the discrepancies between the reported values of the paramagnetic magnetization for \textbf{H}~$\parallel$~\textbf{c}.

\section{Discussion}
\label{conclusion}
To summarize, we have performed magnetometry, transport and small-angle neutron scattering (SANS) measurements on HoNi$_2$B$_2$C single crystals.
Additional SANS intensity $I$ appearing at low temperatures is observed at all fields.
In a low field region $\mu_0 H < 0.58$\,T that includes zero field,
the temperature dependence of the signal is order-parameter-like with a transition at $\approx 5$\,K.
At high fields $\mu_0 H > 1.51$\,T, the SANS increases continuously on cooling across the temperature range probed
(from $\lesssim 3$\,K to $\gtrsim 25$\,K).

At low but finite fields, superconducting vortices will exist in the sample with an arrangement
that is known to be rather disordered from previous local Hall probe~\cite{Dewhurst} and Bitter decoration work~\cite{Vinnikov}.
SANS $I(q)$ profiles similar to those observed here were previously reported on polycrystalline Sr$_{0.9}$La$_{0.1}$CuO$_{2}$
and interpreted as scattering from disordered vortices~\cite{White}.
Such an interpretation cannot account for our HoNi$_2$B$_2$C scattering data since we also observe a SANS signal at zero field.

The agreement between the observed field regimes and the expected metamagnetic transitions,
plus the Porod-law behaviour exhibited in $I(q)$,
lead us to conclude that the low-temperature SANS stems from uncompensated moments at crystallographic grain boundaries and dislocations.
At low fields, these uncompensated moments are associated with domain walls in the antiferromagnetic $\uparrow \downarrow$ phase,
while at high fields they are suggested to stem from paramagnetic moments at the interfaces that align along the field direction
rather than along the nearest [110] easy axis in the bulk ferromagnetic structure.
We emphasize that these magnetic structures form a network of large length scales $D \gg 1300$\,\AA.
Magnetic contrast over similar scales has already been observed in Bitter decoration
of non-superconducting TbNi$_2$B$_2$C and of the normal state of ErNi$_2$B$_2$C~\cite{Vinnikov2003and2009}.
In these high-resolution decoration studies, the lamellar magnetic flux structures observed were thought to be linked to
crystallographic twin boundaries by magnetoelastic stresses.
In HoNi$_2$B$_2$C magnetoelastic effects lead to an orthorhombic distortion
such that the unit cell length along the [110] (tetragonal) direction closest to \textbf{H} shrinks by $\approx 0.2$\,\% compared
to $[\bar{1} 1 0]$~\cite{Kreyssig}.
These large magnetoelastic strains would likely favour the location of magnetic domain walls at crystallographic grain boundaries and dislocations \cite{Cubitt}.
It would be extremely interesting to explore, in future work, the physics of antiferromagnetic magnetoelastic domain wall boundaries
using high-resolution microscopy techniques in HoNi$_2$B$_2$C and in other materials.

\begin{acknowledgements}
We acknowledge the support of the National Institute of Standards and Technology, U.S. Department of Commerce,
in providing the neutron research facilities used in this work.
M.L. acknowledges support from DanScatt.
This work utilized facilities supported in part by the National Science Foundation under Agreement No. DMR-0944772.
\end{acknowledgements}

\end{document}